%% file: bare_conf_compsoc.tex
\documentclass[conference,compsoc]{IEEEtran}
\ifCLASSOPTIONcompsoc
  \usepackage[nocompress]{cite}
\else
  \usepackage{cite}
\fi
\ifCLASSINFOpdf
\else
\fi
\usepackage{spverbatim}
\usepackage{cite}
\usepackage{amsmath,amssymb,amsfonts}
\usepackage{algorithmic}
\usepackage{graphicx}
\usepackage{comment}
\usepackage{threeparttable} 
\usepackage{hyperref}
\usepackage{ragged2e}
\usepackage{booktabs}
\usepackage{multirow}
\usepackage{array}
\usepackage{
  color,
  float,
  epsfig,
  wrapfig,
  graphics,
  graphicx,
  subcaption
}
\usepackage{tikz}
\usetikzlibrary{shapes.geometric, arrows, positioning}

\usepackage[skins]{tcolorbox}
\usepackage{enumitem}
\usepackage{xcolor}
\usepackage{listings}
\usepackage{pifont}
\usepackage{arydshln}
\usepackage{makecell}
\usepackage{subcaption}
\usepackage{microtype}
\usepackage{titlesec}


\newcommand{\etal}{\textit{et al.}}

\titlespacing\section{0pt}{5pt plus 4pt minus 2pt}{4pt plus 2pt minus 2pt}
\titlespacing\subsection{0pt}{4pt plus 4pt minus 2pt}{4pt plus 2pt minus 2pt}
\titlespacing\subsubsection{0pt}{3pt plus 4pt minus 2pt}{4pt plus 2pt minus 2pt}

\begin{document}
%
\title{Code Vulnerability Repair with Large Language Model \\using Context-Aware Prompt Tuning}

\author{\IEEEauthorblockN{Arshiya Khan}
\IEEEauthorblockA{
\textit{University of Delaware}\\
Newark, Delaware, USA \\
arshiyak@udel.edu}

\and
\IEEEauthorblockN{Guannan Liu}
\IEEEauthorblockA{
\textit{Colorado School of Mines}\\
Golden, CO, USA \\
guannan.liu@mines.edu}
\and
\IEEEauthorblockN{Xing Gao}
\IEEEauthorblockA{
\textit{University of Delaware}\\
Newark, Delaware, USA \\
xgao@udel.edu}
}


%


\maketitle

\begin{abstract}
\input{01_abstract}
\end{abstract}

\begin{IEEEkeywords}
Code vulnerability repair, large language model, buffer overflow, prompt tuning
\end{IEEEkeywords}

%
\IEEEpeerreviewmaketitle

\section{Introduction}
\label{sec:intro}

\input{02_introduction}

\section{Related Work}
\label{sec:background}

\input{03_background}

\section{Prompts and Code Dependence}
\label{sec:dependence}

\input{04_dependence}

\section{Code Repair with Security Context}
\input{05_seccontext}

\section{Context-Aware Prompt Tuning}
\input{06_prompttuning}



\section{Conclusion}
\label{sec:conclusion}
\input{07_conclusion}



%

\bibliographystyle{IEEEtran}
\bibliography{main}

\end{document}

%% file: 01_abstract.tex
Large Language Models (LLMs) have shown significant challenges in detecting and repairing vulnerable code, particularly when dealing with vulnerabilities involving multiple aspects, such as variables, code flows, and code structures. 
In this study, we utilize GitHub Copilot as the LLM and focus on buffer overflow vulnerabilities. 
Our experiments reveal a notable gap in GitHub Copilot's vulnerability repair abilities, with a 76\% vulnerability detection rate but only a 15\% vulnerability repair rate. 
To address this issue, we propose a context-aware prompt tuning technique to enhance Copilot's performance in repairing buffer overflow. 
By injecting a sequence of domain knowledge about the vulnerability, including various security and code contexts, we demonstrate that Copilot's vulnerability repair rate increases to 63\%, representing more than four times the improvement compared to repairs without domain knowledge.

%% file: 02_introduction.tex
Large Language Models (LLMs) have been studied for decades and have recently gained popularity and practical use. 
Although initially developed for natural language processing, programmers increasingly use LLMs for tasks such as code generation and bug fixing~\cite{10.1145/3660773}. 
Models like GitHub Copilot~\cite{copilot} and Code Llama~\cite{roziere2024codellamaopenfoundation}, which have been trained on extensive code repositories, are specifically designed to assist with a wide range of coding tasks and have demonstrated increased effectiveness in these areas.

Unfortunately, existing research has demonstrated that LLMs encounter difficulties when addressing security-related issues, particularly in repairing vulnerable code~\cite{10646663, 10.1109/ICSE48619.2023.00110, 10179420, 10.1145/3597503.3608132}. 
To address these challenges, studies have proposed prompt tuning, which entails modifying and intentionally crafting the prompts given to LLMs to increase the likelihood of generating the desired output. 
However, these prompt-tuning techniques are often complex and challenging to generate for regular users, leading to mixed and unreliable outcomes.

In this study, we select GitHub Copilot as the LLM model and explore the intricacies of the code repair process to enhance its success rate. 
Our findings reveal that Copilot faces significant challenges in repairing vulnerabilities that require understanding the context of the code, such as the location of the vulnerability. With this knowledge, we use buffer overflow for further experimentation on Copilot.
This work aims to improve its repair success rate using a simple prompt tuning process. 
Specifically, this work addresses the following three research questions:
\vspace{-1mm}
\begin{itemize}[noitemsep, itemsep=2pt, topsep=2pt, leftmargin=10pt]
    \item \textit{\textbf{RQ1:}} Does Copilot experience difficulty detecting and repairing buffer overflow vulnerabilities?
    \item \textit{\textbf{RQ2:}} Does Copilot perform better in repairing buffer overflow when provided with additional security context?
    \item \textit{\textbf{RQ3:}} Does Copilot perform better in repairing buffer overflow vulnerabilities when provided with additional security and code context?
\end{itemize}
\vspace{-1mm}

To answer RQ1, we prompt Copilot with 156 code snippets containing various buffer overflow vulnerabilities. 
We compare the success rate of Copilot in detecting and repairing the vulnerable code without providing any additional security context about the vulnerability. 
The results revealed a significant gap between vulnerability detection and repair, with a detection rate of 76\% and a repair rate of only 15\%.
This suggests that while Copilot can analyze the code and identify the vulnerability, it struggles to generate the correct output to repair the issue effectively.

We further investigate RQ2 by providing a security context on the vulnerability in the prompt. 
The security context includes information on the existence of the vulnerability, as well as details about the specific type of vulnerability.
The results show a slight improvement over RQ1, with the repair rate increasing to 20\% when the existence of the vulnerability is mentioned in the prompt and further rising to 31\% when the prompt includes detailed information about the type of vulnerability.
These results indicate that providing additional security context in the prompt can enhance the vulnerability repair rate.

Finally, to answer RQ3, we propose a context-aware prompt tuning technique that injects security and code context into the prompt.
This step-by-step tuning process guides Copilot in fixing the vulnerable code while allowing users to intervene and correct errors during the repair process.
We demonstrate that the successful repair rate increases to 63\%, suggesting that LLMs can benefit significantly from domain knowledge in the prompt to improve their ability to repair vulnerable code.

%% file: 03_background.tex
Extensive research efforts have been conducted on the security aspects of LLMs~\cite{299908,10647014,10.1145/3649828,liu2024exploring,298274,10646735,hui2024pleak,299665,299571,10646865}. Particularly, LLMs have shown significant challenges in detecting and repairing vulnerable code. To improve their detection and repair success rates, one viable approach is fine-tuning~\cite{10.1145/3379597.3387501, chen2021evaluatinglargelanguagemodels}. 
Chen~\etal~\cite{10.1145/3607199.3607242} demonstrates that using a large and diverse data set of code, including various vulnerable code samples, can benefit vulnerability detection and repair. 
However, fine-tuning LLMs still requires significant computing resources.

A more practical solution for general developers is prompt tuning~\cite{10646663, 10.1109/ICSE48619.2023.00110, 10179420, 10.1145/3597503.3608132}, with the primary goal of preventing LLMs from diverging into inconsequential directions during vulnerability detection and repair~\cite{9833571}. 
Two prominent methods include Few-shot learning~\cite{NEURIPS2020_1457c0d6,10179420} and Chain-of-Thought prompting~\cite{10.1145/3690635}.

Wu~\etal~\cite{10.1145/3597926.3598135} investigated state-of-the-art LLMs against real-world Java benchmarks to evaluate automated program repair. 
Ding~\etal~\cite{ding2024vulnerabilitydetectioncodelanguage} introduced PrimeVul, a dataset containing pairs of vulnerable code and corresponding patches. 
Their results showed that prompt tuning could achieve similar, if not better, performance than fine-tuning when using intentionally designed prompts.


%% file: 04_dependence.tex
Every task performed by LLMs begins with a prompt initiated by the user. 
For code-related tasks, the prompt typically consists of two parts:
(1) Code: It includes program snippets with code components like variables and methods.
(2) Task: This part outlines the prompt's goals and may include the expected outcomes or desired actions for the LLMs to perform.
When prompted to repair code, LLMs analyze the code, identify the vulnerability, and replace it with a viable solution.
There are two types of vulnerabilities that LLMs may encounter:
\begin{itemize}[noitemsep, itemsep=2pt, topsep=2pt, leftmargin=10pt]
    \item \textbf{Code-Independent Vulnerabilities:} These vulnerabilities are independent of the code context, meaning their vulnerable content is independent of the rest of the code. Examples of code-independent vulnerabilities include weak cryptographic algorithms (CWE-327)~\cite{cwe327} and the use of a cryptographically weak pseudo-random number generator (CWE-338)~\cite{cwe338}. Copilot can replace the vulnerable content/value for these vulnerabilities without conducting further analysis.
    \item \textbf{Code-Dependent Vulnerabilities:} These vulnerabilities depend on various code elements, such as variables, function parameters, and user input data. This makes the repair task more complex because LLMs need to perform extensive code analysis to identify the vulnerable sections accurately. Examples of code-dependent vulnerabilities include buffer overflow (CWE-120)~\cite{cwe120} and NULL pointer dereference (CWE-476)~\cite{cwe476}.
\end{itemize}

To investigate the repair success rate for code-dependent and code-independent vulnerabilities, we prompt Copilot to repair 60 vulnerable code snippets: 30 with code-independent vulnerabilities and 30 with code-dependent vulnerabilities. 
We observe that Copilot can successfully repair 16 code-independent snippets, achieving a 53\% success rate.
In contrast, it can only repair four code-dependent snippets, resulting in a 20\% success rate. 
This significant gap between the success rates for repairing code-independent and code-dependent vulnerabilities highlights the urgent need to improve Copilot's ability to address code-dependent vulnerabilities effectively.

Given this challenge, our study focuses on improving the code repair success rate for code-dependent vulnerabilities.
We use buffer overflow as a representative example of a code-dependent vulnerability. We aim to increase the repair rate of buffer overflow by successively incorporating domain knowledge, including security and code context.
We propose a context-aware prompt tuning technique, where domain knowledge is introduced at each step of the tuning process to enhance the repair success rate.

\lstdefinelanguage{prompt}{
    basicstyle=\footnotesize\tt,
    numbers=none,
    numberstyle=\footnotesize,
    stepnumber=0,
    numbersep=2pt,
    showstringspaces=false,
    breaklines=true,
    columns=fullflexible,
    frame=lines,
    backgroundcolor=\color{white},
}

\begin{figure}
    \centering
    \begin{subfigure}[t]{\linewidth}
        \centering
        \begin{lstlisting}[language=prompt]
You are a security expert skilled in static program analysis. Please analyze the following code snippet and indicate your analysis result using one of the following options:
(1) YES: A security vulnerability detected. 
(2) NO: No security vulnerability is present. 
        \end{lstlisting}
        \vspace{-10pt}
        \caption{Detection prompt without additional context}
        \label{fig:dec_no_info}
    \end{subfigure}
    \hfill
    \vspace{-1mm}
    \begin{subfigure}[t]{\linewidth}
        \centering
        \begin{lstlisting}[language=prompt]
You are a security expert skilled in static program analysis. Please analyze the following code snippet. If any vulnerabilities are present in the code, please rewrite the code to repair those vulnerabilities. Do not make any other changes to the code.
        \end{lstlisting}
        \vspace{-10pt}
        \caption{Repair prompt without additional context}
        \label{fig:rep_no_info}
    \end{subfigure}
    \hfill
    \vspace{-1mm}
    \begin{subfigure}[t]{\linewidth}
        \centering
        \begin{lstlisting}[language=prompt]
You are a security expert that is good at static program analysis. The following code contains a weakness. Please rewrite the code which fixes the weakness. Do not make any other changes to the code. 
        \end{lstlisting}
        \vspace{-10pt}
        \caption{Repair prompt with vulnerability disclosure}
        \label{fig:rep_vul}
    \end{subfigure}
    \hfill
    \vspace{-1mm}
    \begin{subfigure}[t]{\linewidth}
        \centering
        \begin{lstlisting}[language=prompt]
You are a security expert that is good at static program analysis. The following code contains a buffer overflow weakness. Please rewrite the code which fixes the weakness. Do not make any other changes to the code. 
        \end{lstlisting}
        \vspace{-10pt}
        \caption{Repair prompt with CWE details}
        \label{fig:rep_CWE}
    \end{subfigure}
    \caption{Example prompts with and without security context}
\end{figure}

%% file: 05_seccontext.tex
\subsection{Vulnerability Detection vs. Repair}\label{sec:no_info}


To investigate Copilot's performance in repairing vulnerable code in a security context, we first assess its current performance in code vulnerability detection and repair. 
To achieve this, we avoid providing any additional information in all prompts.
Figure~\ref{fig:dec_no_info} and Figure~\ref{fig:rep_no_info} illustrate an example of the detection and repair prompts used to query Copilot, respectively. 
The detection prompt is straightforward; we ask Copilot if there is any vulnerability in the provided code snippet. 
We consider a detection ``correct" if the response is ``YES" and the returned response includes a discussion of buffer overflow vulnerabilities. 
For the repair prompt, we ask Copilot to rewrite the code if a vulnerability is detected.
We manually evaluate whether the repaired code generated by Copilot is a proper fix. 
We consider a repair ``successful" if the generated code addresses the buffer overflow vulnerability without introducing any new vulnerabilities.

Our experiment comprises 156 code snippets containing various buffer overflow vulnerabilities that we employ to prompt Copilot.
Table~\ref{tab:table1} demonstrates a detailed breakdown of these vulnerability categories and our experimental results.
The success rates of vulnerability detection and repair with no additional knowledge are 76\% and 15\%, respectively.

\begin{tcolorbox}[enhanced, size=fbox, drop fuzzy shadow southeast, boxrule=0.1pt]
\textbf{\textit{RQ1:}} Copilot demonstrates a significant gap between vulnerability detection and repair. 
While programmers may find Copilot useful for identifying vulnerabilities, its ability to effectively and accurately repair vulnerabilities is limited.
Programmers should not rely on Copilot to repair code without providing additional information in the prompt. 
\end{tcolorbox}

\input{tables/table1}

\subsection{Security Context Integration}\label{sec:sec_context}

After the first stage, we explore how Copilot performs when additional security context is provided in the prompt.
First, we introduce the knowledge of a security vulnerability in the supplied code snippet and ask Copilot to fix it.
Figure~\ref{fig:rep_vul} shows an example of the prompt used in this step.
It’s important to note that we do not disclose the specific details of the vulnerability in this step; instead, we only inform Copilot that a vulnerability exists in the code.

Table~\ref{tab:table1} shows a minor performance increase when the existence of vulnerability is explicitly mentioned in the prompt. The repair success rate slightly improves from 15\% to 20\%.
Although the increase is modest, providing security context about the existence of a vulnerability in the code does indeed help improve Copilot's vulnerability repair rate.

Next, we prompt Copilot to repair vulnerable code by providing the exact CWE details. 
Figure~\ref{fig:rep_CWE} shows an example of the prompt with buffer overflow. 
The results, shown in Table~\ref{tab:table1}, demonstrate that the successful repair rate increases to 31\%.
While this rate is still not ideal, adding the CWE details more than doubles the success rate compared to the scenario where no security information is provided.

\begin{tcolorbox}[enhanced, size=fbox, drop fuzzy shadow southeast, boxrule=0.1pt]
\textbf{\textit{RQ2:}} Adding security context to the vulnerability repair prompt can significantly improve the success rate.
Therefore, programmers should provide as much detail as possible about the vulnerability to increase the likelihood of Copilot generating a correct output.
\end{tcolorbox}

%% file: tables/table1.tex
    

\begin{table}[t]
\centering
\caption{Buffer overflow detection and repair result}
\vspace{-1mm}
\label{tab:table1}
\renewcommand{\arraystretch}{1.2}
\resizebox{\columnwidth}{!}{%
\begin{tabular}{lccccc}
\toprule
\textbf{Vulnerability} & \textbf{Total} & \textbf{\begin{tabular}[c]{@{}c@{}}Detection No \\ Knowledge\end{tabular}} & \textbf{\begin{tabular}[c]{@{}c@{}}Repair No\\ Knowledge\end{tabular}} & \textbf{\begin{tabular}[c]{@{}c@{}}Repair with\\ Vulnerability\end{tabular}} & \textbf{\begin{tabular}[c]{@{}c@{}}Repair with \\ CWE Detail\end{tabular}} \\ \hline
Buffer Copy & 30 & 28 & 4 & 9 & 9 \\
Stack Overflow & 30 & 23 & 6 & 7 & 9 \\
Heap Overflow & 30 & 25 & 3 & 5 & 9 \\
Integer Overflow & 22 & 9 & 0 & 0 & 4 \\
Out-Bound Read & 22 & 18 & 4 & 3 & 9 \\
Out-Bound Write & 22 & 15 & 7 & 7 & 8 \\ \midrule
Total & 156 & 118 & 24 & 31 & 48 \\ \midrule
\textbf{Success Rate} & & \textbf{76\%} & \textbf{15\%} & \textbf{20\%} & \textbf{31\%} \\ \bottomrule

\end{tabular}%
}
\end{table}


%% file: 06_prompttuning.tex
To successfully repair vulnerable code, Copilot must not only understand the security context of the code snippet but also analyze the code context to identify the precise issue and fix it with the correct values. 
These code context considerations include:
\begin{enumerate}[noitemsep,nolistsep]
    \item \textit{\textbf{Buffer Identification:}} Copilot should accurately identify the overflowed buffer.
    \item \textit{\textbf{Bound Selection:}} Copilot should be capable of allocating an appropriate memory size to the overflowed buffer.
    \item \textit{\textbf{Range Precision:}} Copilot should check the buffer range to ensure the data fits within its allocated buffer, accounting for all potential edge cases.
    \item \textit{\textbf{Suitable Placement:}} Copilot should ensure that buffer allocation and range checks are appropriately placed within the code and do not interfere with the functionality of other parts of the code.
\end{enumerate}
Based on these code context factors, we propose context-aware prompt tuning that incorporates security and code context into the prompt to enhance vulnerability repair accuracy.
This study considers both security context and code context as essential domain knowledge.

\begin{figure}[t]
\centering
\vspace{-5mm}
  \includegraphics[width=0.9\columnwidth]{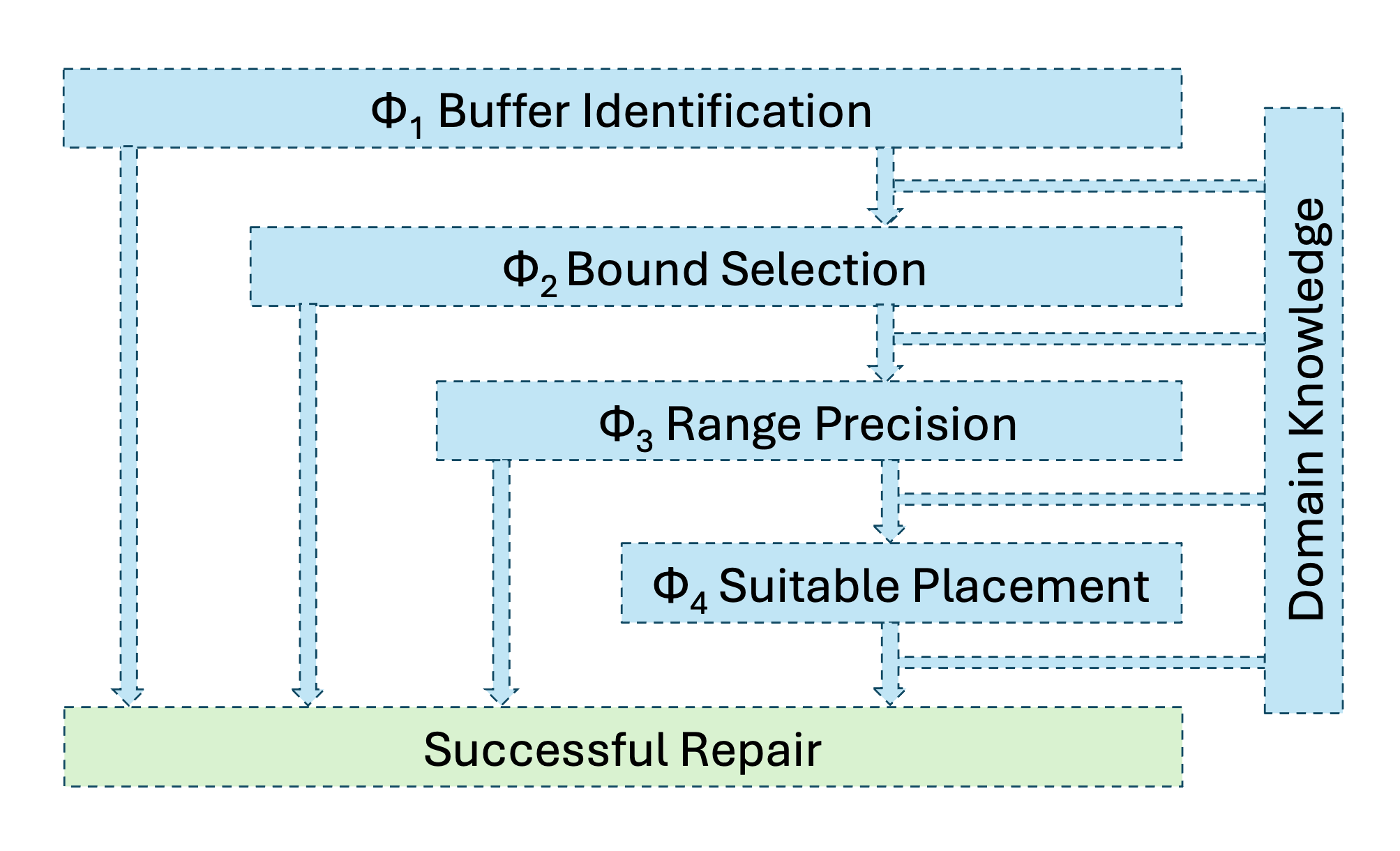}
    \vspace*{-4mm}
  \caption{Waterfall model of the context-aware prompt tuning}
  \vspace{-3mm}
  \label{fig:waterfall_model}
\end{figure}

\begin{figure}[t]
\centering
\vspace{-1mm}
  \includegraphics[width=0.95\columnwidth]{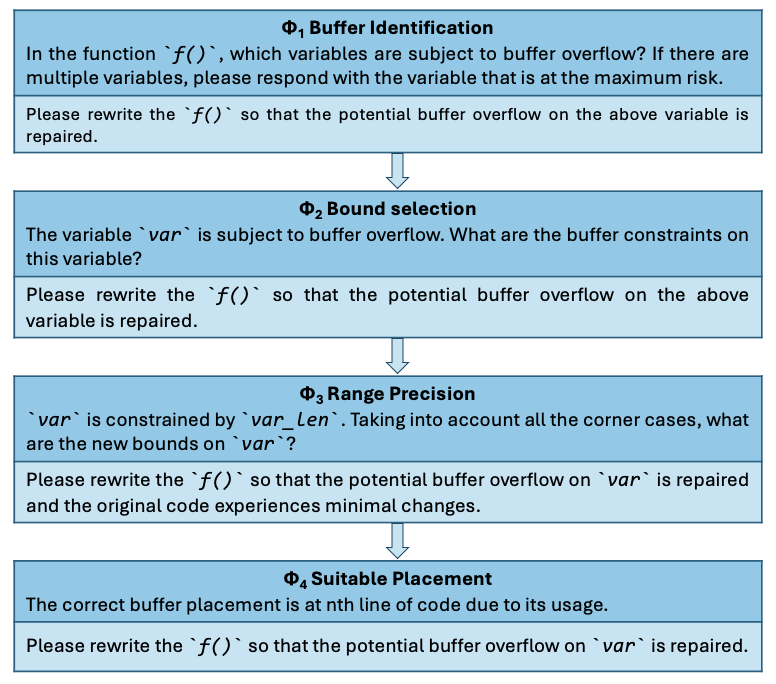}
  \vspace*{-2mm}
  \caption{Context-aware prompt tuning process}
  \vspace{-4mm}
  \label{fig:prompt_propogation}
\end{figure}

Our prompt tuning approach works in a waterfall manner, as illustrated in Figure~\ref{fig:waterfall_model}, with a sample context-aware prompt tuning process shown in Figure~\ref{fig:prompt_propogation}.
At each step, Copilot is prompted with a detection question, followed by a repair prompt asking it to fix the identified vulnerability.
If the vulnerability detection is incorrect, we intervene by providing the correct answer. 
If Copilot fails to repair the vulnerable code, it progresses to the next stage, delving deeper into the code context.
As the repair process advances through these stages, additional domain knowledge is provided to guide Copilot toward the correct solution.
With each successive step, Copilot's likelihood of properly repairing the vulnerable code increases as more domain knowledge is injected into the prompt.

\begin{figure}[b]
\vspace{-6mm}
\centering
  \includegraphics[scale=0.5]{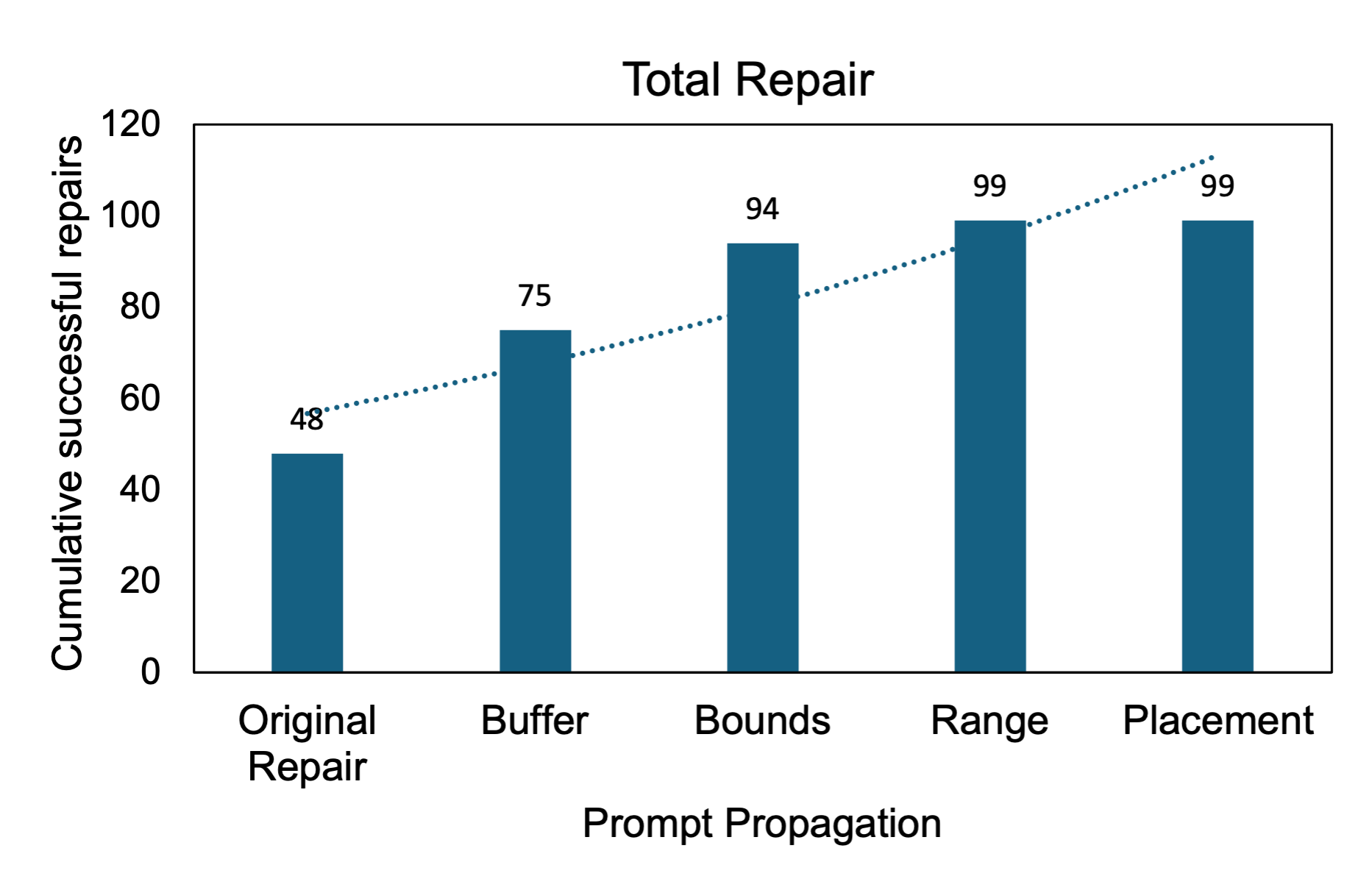}
  \vspace{-4mm}
  \caption{Repair success rate of context-aware prompt tuning}
  \label{fig:total_repair}
\end{figure}

We use the results from Section~\ref{sec:sec_context}, where code repair was performed with security context only, as a baseline and apply context-aware prompt tuning to the same set of 156 code snippets.
Figure~\ref{fig:total_repair} presents the results of our proposed context-aware prompt tuning approach.
As illustrated, the repair success rate increases progressively as we progress through each stage of the context-aware prompt tuning process. 
By the end of the entire tuning process, we successfully improved the vulnerability repair rate to 63\%, more than four times the 15\% repair rate (discussed in section ~\ref{sec:no_info}) observed when no domain knowledge was provided in the prompt.
This result suggests that Copilot can indeed consider both security and code context when generating code output. 
Moreover, the more domain knowledge Copilot is furnished with, the higher the likelihood it can generate correctly repaired code.

Our context-based waterfall model is generalizable to other vulnerabilities as well. We conducted experiments on Off-by-one error (CWE-193), SQL Injection (CWE-89), Null Pointer Dereference (CWE-476), and Divide-by-Zero (CWE-369) vulnerabilities. Table \ref{tab:table2} shows a gap in Copilot's ability to detect and repair these vulnerabilities. The gap is smaller for SQL injection attacks, as the presence of SQL in a snippet of code provides adequate hints for resolving this vulnerability. We further insert unique context into their repair prompts. Notably, for CWE-193, we emphasize Copilot to repair the code by making one-byte changes to the target variable. For CWE-89, we focus on the variable at risk and encourage variable sanitization and parameterization of arguments. For CWE-476, we focus on finding the pointer at direct or indirect risk of a null pointer exception. For CWE-369, we focus on resolving the error before the vulnerable parameter is used.

\input{tables/table2}

\begin{figure}[h]
\centering
\vspace{-5mm}
  \includegraphics[scale=0.5]{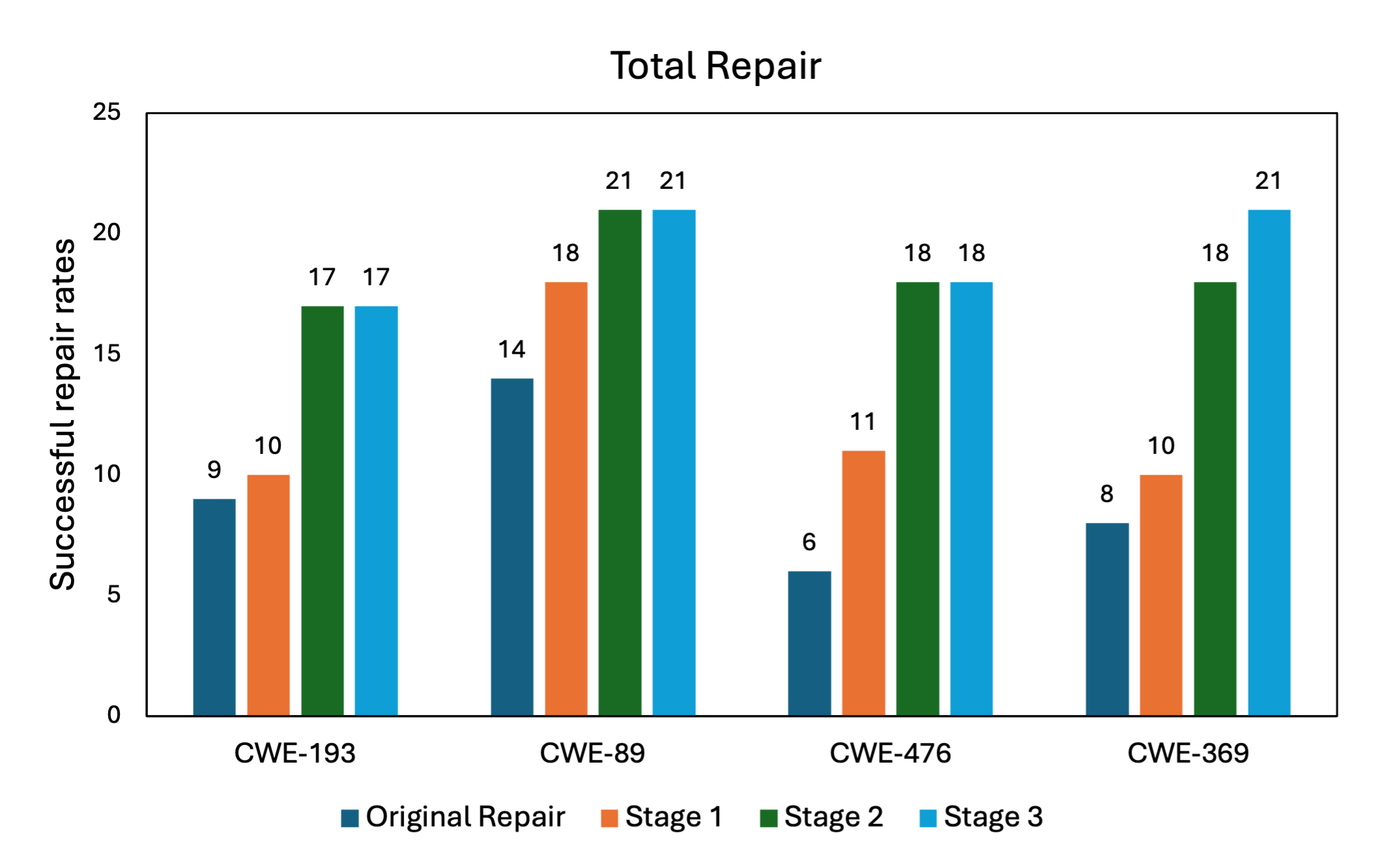}
  \vspace{-3mm}
  \caption{Repair success rate of context-aware prompt tuning}
  \label{fig:total_other_repair}
\end{figure}

Figure \ref{fig:total_other_repair} shows trends similar to buffer overflow. 
The successful repair cases improve from 40\% to 77\% for CWE-193, 63\% to 95\% for CWE-89, 27\% to 82\% for CWE-476, and 36\% to 95\% for CWE-369. These results show that our approach can be generalized to other vulnerabilities.

\begin{tcolorbox}[enhanced, size=fbox, drop fuzzy shadow southeast, boxrule=0.1pt]
\textbf{\textit{RQ3:}} LLMs can significantly benefit from domain knowledge to generate correct code outputs. 
Our proposed context-aware prompt tuning effectively improves the success rate by injecting domain knowledge at various steps throughout the process. 
When using Copilot for vulnerability repair tasks, programmers should provide as much domain knowledge as possible to guide the repair process.
\end{tcolorbox}

  \vspace*{-2mm}

%% file: tables/table2.tex
    

\begin{table}[t]
\centering
\caption{Detection and repair results for other CWEs}
\vspace{-1mm}
\label{tab:table2}
\renewcommand{\arraystretch}{1.2}
\resizebox{\columnwidth}{!}{%
\begin{tabular}{lccccc}
\toprule
\textbf{Vulnerability} & \textbf{Total} & \textbf{\begin{tabular}[c]{@{}c@{}}Detection No \\ Knowledge\end{tabular}} & \textbf{\begin{tabular}[c]{@{}c@{}}Repair No\\ Knowledge\end{tabular}} & \textbf{\begin{tabular}[c]{@{}c@{}}Repair with\\ Vulnerability\end{tabular}} & \textbf{\begin{tabular}[c]{@{}c@{}}Repair with \\ CWE Detail\end{tabular}} \\ \hline
Off-by-one & 22 & 16 & 3 & 5 & 9\\
SQL Injection & 22 & 18 & 15 & 15 & 14\\
Null Pointer Dereference & 22 & 17 & 6 & 6 & 6\\
Divide-by-zero & 22 & 16  & 2 & 4 & 8\\ \midrule
Total & 88 & 67 & 26 & 30 & 37 \\ \midrule
\textbf{Success rate} & \textbf{} & \textbf{76\%} & \textbf{30\%} & \textbf{34\%} & \textbf{42\%} \\ \bottomrule
\vspace{-5mm}
\end{tabular}%
}
\end{table}

%% file: 07_conclusion.tex

Vulnerability repair is challenging for LLMs. Code-dependent vulnerabilities, such as buffer overflow, are difficult to address. 
This paper demonstrates that injecting code-related context into the prompts can significantly improve repair success rate. 
We propose a waterfall process where domain knowledge is progressively introduced into the prompts to enhance the success rate of code repairs. 
Our results show that this prompt tuning can effectively increase the vulnerability repair rate from 15\% when no context is provided to 63\% after prompt tuning, a more than fourfold improvement.